\documentclass[10pt]{iopart}
\usepackage{amssymb,graphicx}
\usepackage{bm,float}
\usepackage{xcolor,cite,doi}
\usepackage{hyperref}
\hypersetup{colorlinks=true, linkcolor=blue!80!black, urlcolor=blue!80!black, citecolor=blue!80!black}

\newcommand{\exb}{{\mathbf{E} \hskip -1pt \times \hskip -1pt \mathbf{B}}}

\newcommand\bunit{B_\mathrm{unit}}

\begin{document}

\title{Dynamic Optimization Workflow for Gyrokinetic Stability and Transport}

\author{E.A.~Belli$^{1}$ \footnote{Author to whom any correspondence should be addressed.}, J.~Larson$^{2}$,  J.~Candy$^1$, I.~Sfiligoi$^3$}
\address{$^1$ General Atomics, P.O.~Box 85608, San Diego, CA 92186-5608, USA}
\address{$^{2}$ Argonne National Laboratory, 9700 S. Cass Ave., Lemont, IL 60439}
\address{$^3$ University of California San Diego, La Jolla, CA 92093 USA}
\ead{bellie@fusion.gat.com}

\begin{abstract}
  A dynamic optimization workflow for linear gyrokinetic stability that couples the gyrokinetic code CGYRO with the libEnsemble framework for ensemble management has been developed. The workflow is designed to efficiently explore multi-dimensional plasma parameter spaces and to address the practical challenges of simulation-driven optimization in fusion plasma physics, including expensive objective evaluations, unavailable gradient information, non-smoothness arising from transitions between dominant instability branches, and simultaneous multi-core simulations on high-performance computing resources.   Rather than optimizing the growth rate at a single wavenumber, the objective maximizes the growth rate-to-wavenumber ratio over the ion-scale spectrum.  A derivative-free, model-based optimization algorithm employs manifold sampling to handle the nonsmooth composite structure of the objective function.  The approach is applied to the optimization of local magnetic flux-surface shaping parameters, with an equilibrium baseline drawn from DIII-D negative triangularity experimental parameters. The workflow is general and extensible and provides a foundation for future optimization studies involving nonlinear transport objectives and higher-dimensional parameter spaces.
\end{abstract}

\section{Introduction}
\label{sec.intro}

Design of fusion reactors and fusion scenarios requires new workflows capable of efficiently exploring large multi-dimensional parameter spaces.  Systematic optimization of gyrokinetic turbulence and stability is a central component for confinement and performance prediction.  However, even linear gyrokinetic simulations require multi-core high-performance computing resources.  Thus, for optimization studies, adaptive automated sampling strategies are needed to coordinate concurrent simulations across large-scale computational resources.

In this work, we present a dynamic optimization workflow that integrates the gyrokinetic code CGYRO~\cite{candy:2016,candy:2019,sfiligoi:2023} within the libEnsemble framework~\cite{Hudson2023,libEnsembleManual} for ensemble management to facilitate solution of parameter optimization problems for linear gyrokinetic stability.  The framework is designed to be general.  It accommodates an objective function based on the gyrokinetic linear growth rate that can be defined over any subset of the local equilibrium parameter space.

To demonstrate the workflow, we apply it to the optimization of magnetic equilibrium flux-surface shaping parameters.  Magnetic equilibrium geometry in tokamak plasmas has long been known as a key influencing factor on plasma stability~\cite{troyon:1984}, confinement~\cite{iter:1999b,doyle:2007}, and turbulent transport~\cite{rewoldt:1982,waltz:1999,kendl:2006,kinsey:2007,belli:2008b,angelino:2009}. The shaping parameter space is intrinsically multi-dimensional and provides the basis for a controlled study of the impact on drift-wave instabilities.  The libEnsemble-CGYRO workflow adaptively samples this landscape, identifying shaping configurations that optimize linear stability objectives, and demonstrates the broader potential of the framework for gyrokinetic parameter optimization.

The remainder of this paper is organized as follows. The numerical optimization algorithm is described in Sec.~\ref{sec.alg}, with the choice of objective function in Sec.~\ref{sec.obj} and details of the libEnsemble-CGYRO workflow in Sec.~\ref{sec.libens}.  Sec.~\ref{sec.shaping} describes the local shaping parameterization and DIII-D equilibrium baseline for the optimization study. The results for multi-dimensional shaping optimization are shown in Sec.~\ref{sec.results}.   Finally, a summary and  discussion of the broader application are given in Sec.~\ref{sec.summary}.

\section{Optimization algorithm}
\label{sec.alg}

Numerical optimization problems are commonly formulated in terms of a vector of decision variables $x \in \mathbb{R}^n$ and a scalar objective function $f(x)$ that measures the quantity of interest. 
For example, the variables $x$ could represent parameters that define a local magnetic equilibrium configuration, while the objective function $f$ measures the gyrokinetic stability. 
Classical optimization methods often assume that gradient information $\nabla f(x)$ is available or can be computed efficiently.
For many problems in fusion and gyrokinetics, however, the objective evaluations require expensive plasma simulations and analytic derivative information that is either unavailable or prohibitively costly to obtain. 
In addition, simulation outputs may exhibit noise, non-smoothness, or changes in the dominant instability branch, further complicating the use of gradient-based approaches.

Optimization methods that do not require derivative information are generally referred to as derivative-free optimization (DFO) methods~\cite{LMW2019}.
Among the most widely used DFO approaches are evolutionary and population-based methods, which repeatedly sample and compare large collections of candidate solutions.
While robust, such methods can require hundreds or thousands of objective evaluations even during initialization or early exploration of the parameter space.
When a single gyrokinetic evaluation may require minutes or hours on high-performance computing resources, this evaluation cost becomes prohibitive.

For expensive simulation-driven optimization problems, model-based DFO methods are often more effective because they seek to reduce the total number of objective evaluations.
These approaches maintain and reuse information from all previously evaluated parameter vectors, including objective values and available additional intermediate simulation quantities.
Local surrogate models are constructed and updated throughout the optimization process, and these models are repeatedly minimized to propose new candidate iterates.
Although the associated linear algebra and model-management costs can be significant relative to simpler heuristic methods, such overhead is negligible compared to the cost of large-scale plasma simulations.
Incurring additional algorithmic work to reduce the number of expensive gyrokinetic evaluations is highly advantageous in this regime.

For this work, the optimization algorithm comes from the 
IBCDFO package, using a manifold-sampling method for non-smooth
composite optimization~\cite{KLW2018,Larson2022,ibcdfo}.
Manifold sampling is
designed for objectives of the form $h(F(x))$, where $F(x)$ is a smooth
vector-valued map obtained from expensive simulations and $h(\cdot)$ is a
non-smooth selection operator.
Rather than approximating the
full non-smooth objective $f$ directly, the manifold sampling constructs local models of the
smooth components of $F$ and combines them using the known form of $h$ to (hopefully) achieve a more accurate surrogate
of the objective.

\section{Objective function}
\label{sec.obj}

In this work, the smooth components $F_i(x)$ are derived from linear growth rates $\gamma$ of drift-wave instabilities computed by the gyrokinetic code CGYRO.  For the scalar objective function, rather than minimizing the linear growth rate at a single binormal wavenumber $k_\theta$, we consider the maximum value of the ratio $\hat{\gamma}/\hat{k}$ over the low-wavenumber (ion-scale) spectrum in the range $0 < \hat{k} \le 0.75$.  Here the linear growth rate is normalized as $\hat{\gamma} = \gamma \; (a/c_s)$, where $a$ is the mid-plane minor radius and $c_s=\sqrt{T_e/m_D}$ is the sound speed, where $T_e$ is the electron temperature and $m_D$ is the deuteron mass.  The binormal wavenumber is normalized as $\hat{k} = k_\theta \rho_s$, where $\rho_s=c_s/\Omega_s$ is the sound gyroradius and $\Omega_s = e \bunit /(m_D c)$ with
$\bunit(r)=(q/r)(d\psi/dr)$, defined with respect to $q$ (safety factor), and $\psi$ (poloidal flux divided by $2\pi$)~\cite{candy:2009b}.  Thus, the smooth components are the mode growth-rate outputs
$\hat{\gamma}(x_n,\hat{k}_{j})$ evaluated over linearly spaced values of $\hat{k}$, where $x_n$ is representative of the multi-dimensional optimization parameters (e.g. shaping parameters that will be defined in Sec.~\ref{sec.shaping}), and the scalar objective can be written as
\begin{equation}
G(x_n) \doteq \max_j \frac{\hat{\gamma}(x_n,\hat{k}_{j})}{\hat{k}_{j}} \; .
\label{eqn_obj}
\end{equation}
 The significance of the ratio $\hat{\gamma}/\hat{k}$ is that it is related to the zonal root mean square $\exb$ velocity under the hypothesis of a zonal $k_x$ (radial wavenumber) mixing saturation mechanism for the nonlinear turbulence~\cite{staebler:2017}.  Thus, it is assumed that this ratio may be better correlated with the nonlinear transport than $\gamma$ at a single $k_\theta$.  
Optimization seeks parameter settings that minimize the most unstable
normalized mode across all sampled values of $\hat{k}$, while naturally handling the switching behavior that occurs when different $\hat{k}$ modes become dominant. 

An alternative to derivative-free optimization is the adjoint method.  An adjoint method computes the gradient of a scalar objective with respect to all input parameters simultaneously by solving an auxiliary system derived from governing equations, rather than by perturbing each parameter individually.  This approach has been used previously for linear gyrokinetic stability optimization \cite{acton:2024}, differentiable gyrokinetic turbulence simulation via automatic differentiation \cite{artigues:2026} (which is mathematically equivalent to the continuous adjoint method), and shape optimization of stellarator coils \cite{paul:2018}.  In the case of Ref.~\cite{acton:2024}, gradients of the linear growth rate with respect to shaping parameters were efficiently calculated for a gyrokinetic stability objective similar to the one considered here, though optimizing the growth rate at a single binormal wavenumber.   The adjoint approach has the significant advantage that the computational cost of evaluating the gradient is independent of the dimension of the parameter space. However, the method requires analytic derivation and numerical implementation of adjoint equations within the gyrokinetic code (or, in the case of automatic differentiation, a fully differentiable implementation of the simulation itself). In contrast, the derivative-free approach adopted here requires no modification to CGYRO and naturally accommodates the non-smooth composite objective arising from transitions between dominant instability branches across the wavenumber spectrum. The adjoint method is thus advantageous when gradient information is available and the objective is smooth, while the manifold-sampling DFO method is better suited to non-smooth objectives defined over a spectrum of wavenumbers.

\section{libEnsemble workflow}
\label{sec.libens}

The numerical optimization workflow is implemented within libEnsemble~\cite{Hudson2023,libEnsembleManual}, a scalable coordination framework designed for managing large ensembles of simulations on parallel computing systems.
libEnsemble use a worker--manager model, where the central manager process allocates computational
resources, launches simulation evaluations, receives returned results, and
dynamically redistributes work among workers.
Such a workflow is particularly
well-suited to expensive plasma physics calculations because simulation jobs
may have heterogeneous runtimes or fail to return in a timely manner.
In such
cases, libEnsemble can detect unresponsive evaluations, reclaim
allocated resources, and redeploy those resources to productive tasks.  This
maintains high machine utilization. The
decision logic governing which parameter vectors are evaluated next is
a user-defined generator function, which allows
optimization algorithms to be embedded directly within the
ensemble workflow.

\section{Shaping parameterization and DIII-D baseline equilibrium}
\label{sec.shaping}

For the shaping optimization application analysis, we use the local magnetohydrodynamic (MHD) equilibrium model known as the Miller extended harmonic (MXH) parameterization~\cite{arbon:2021}, which is based on the following specification for a closed contour:
\begin{eqnarray}
R(r,\theta) & = R_0(r) + r \cos \theta_R \; , \\
Z(r,\theta) & = Z_0(r) + \kappa r \sin \theta \; ,
\end{eqnarray}
where
\begin{equation}
\theta_R \doteq \theta + c_0(r) + \sum_{n=1}^{N} \left[ c_n(r) \cos (n\theta  )+ s_n(r) \sin (n\theta) \right] \; .
\end{equation}
Here the shaping parameters physically represent the major radius $R_0$, elevation $Z_0$, elongation $\kappa$, up-down asymmetric parameters $c_n$, and up-down symmetric parameters $s_n$.  The parameter $s_1$ is related to the triangularity $\delta$ ($s_1 = \sin^{-1} \delta$), and $s_2$ is related to the squareness $\zeta$ ($s_2=-\zeta$).  The radial derivatives (shear) of these parameters are also required.  For the harmonic coefficients, we define these as $s_{c_n} \doteq r \partial c_n/\partial r$ and $s_{s_n} \doteq r \partial s_n/\partial r$ (such that $s_{s_1} = s_\delta$ and $s_{s_2} = s_\zeta$).  The radial derivatives of $R_0$, $Z_0$, and $\kappa$ are defined as $dR_0/dr$, $dZ_0/dr$, and $s_\kappa = (r/\kappa) \partial \kappa/\partial r$, respectively.

In physical experiments, the shaping parameters are not fully independent and tend to be correlated with each other, due to consistency with the global equilibrium. However, the local equilibrium model allows for varying the individual shaping parameters independently at a single flux surface, enabling a controlled study of their impact on gyrokinetic stability.  These parameters form the multi-dimensional search space over which the workflow operates.

We consider an equilibrium baseline drawn from DIII-D negative-triangularity (NT) experimental discharge \#171646, in the electron cyclotron heated (ECH)-only phase at $t=2.8ms$~\cite{mcclenaghan:2024}.  Negative triangularity is a promising configuration for reactors for yielding high levels of confinement without edge-localized modes (ELMs)~\cite{austin:2019,coda:2021}. We choose one radial zone ($r/a=0.75$) as a representative shaped flux surface and artificially vary the shaping parameters using the MXH formulation. This allows us to study the gyrokinetic landscape across a broad range of magnetic geometries, while maintaining consistency with experimentally relevant conditions.  The full local equilibrium parameters for this case are given in Table \ref{table_equil}.

While switching plasma shape from NT to positive triangularity (PT) can indirectly affect MHD activity and profile evolution, here we focus only on the direct impact of shaping parameters.    All equilibrium and geometry parameters are fixed except for $\delta$ and $\zeta$ and their radial derivatives. The latter, as specified, are either artificially suppressed ($s_\delta=s_\zeta=0$), or varied as follows via linear interpolation, anchored to the experimental shear at the experimental shaping values and vanishing in the limit of an ellipse:
\begin{eqnarray}
s_\delta(\delta)  & = \left(s_\delta \right)_{exp} \frac{\delta}{\delta_{exp}} \; ; 
\label{eqn_sdelta} \\
s_\zeta(\zeta) & = \left(s_\zeta \right)_{exp} \frac{\zeta}{\zeta_{exp}} \; ,
\label{eqn_szeta}
\end{eqnarray}
where the experimental values are defined in Table \ref{table_equil}.
The elongation $\kappa$ is fixed and thus not included in the optimization parameters since it is well-known to have a strong stabilizing effect in tokamaks and thus tends to drive the results toward the upper bound of any search domain.

\begin{table}[t]
\begin{center}
\resizebox{\textwidth}{!}{%
\begin{tabular}{|c|c|c||c|c|c|}
\hline
Parameter & Description & DIII-D Value & Parameter & Description & DIII-D Value  \\
\hline
$r/a$ & minor radius & 0.75 & $T_i/T_e$ & ion temperature & 1.0 \\
$n_i/n_e$ & ion density & 1.0  & $a/L_{na}$ & $-(d/dx) \ln n_a$ & 1.71 \\
$a/L_{Te}$ & $-(d/dx) \ln T_e$ & 4.36 &  $a/L_{Ti}$ & $-(d/dx)\ln T_i$ &  2.46 \\
$\beta_{e,{\rm unit}}$ & $ 8\pi n_e T_e/B_{\rm unit}^2$ & 2.73 $\times 10^{-4}$  & 
$(a/c_s) \nu_{ee} $ & electron collision rate &  0.25  \\
$q$      & safety factor & 2.168 & $s$       & magnetic shear  & 1.847 \\
$R_0/a$  & major radius  & 2.812 & $dR_0/dr$ & Shafranov shift & -0.280 \\
$Z_0/a$ & elevation  & -0.0137 & $dZ_0/dr$ & elevation gradient & -0.00215 \\
$\kappa$ & elongation & 1.252 & $s_\kappa$ & $(x/\kappa) (d\kappa/dx)$ & 0.151 \\
$\delta$ & triangularity & -0.194 & $s_\delta$ &  $x (d\delta/dx)$ & -0.479 \\
$\zeta$ & squareness & -0.0738 & $s_\zeta$ & $x (d\zeta/dx)$ & -0.309 \\
\hline
\end{tabular}
}
\caption{Local equilibrium profile and baseline geometry parameters for DIII-D \#171646 at $t=2.8ms$.  Here $x \doteq r/a$. Note that $\beta_{e,{\rm unit}}$ was reduced by 1/3 from its experimental value.}
\end{center}
\label{table_equil}
\end{table}

\section{Shaping optimization results}
\label{sec.results}

\begin{figure}
    \centering
    \includegraphics[width=0.49\textwidth]{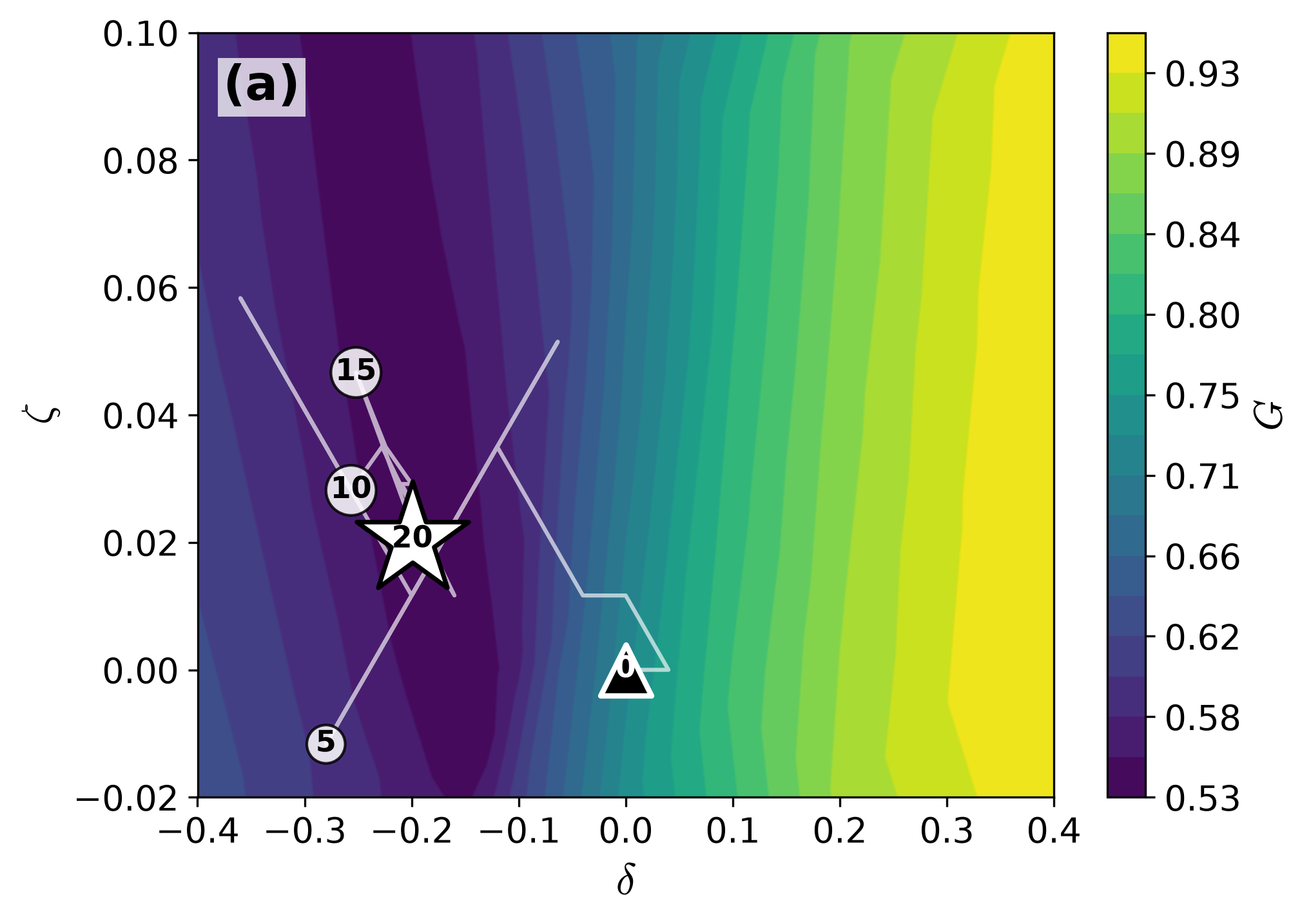}
       \includegraphics[width=0.49\textwidth]{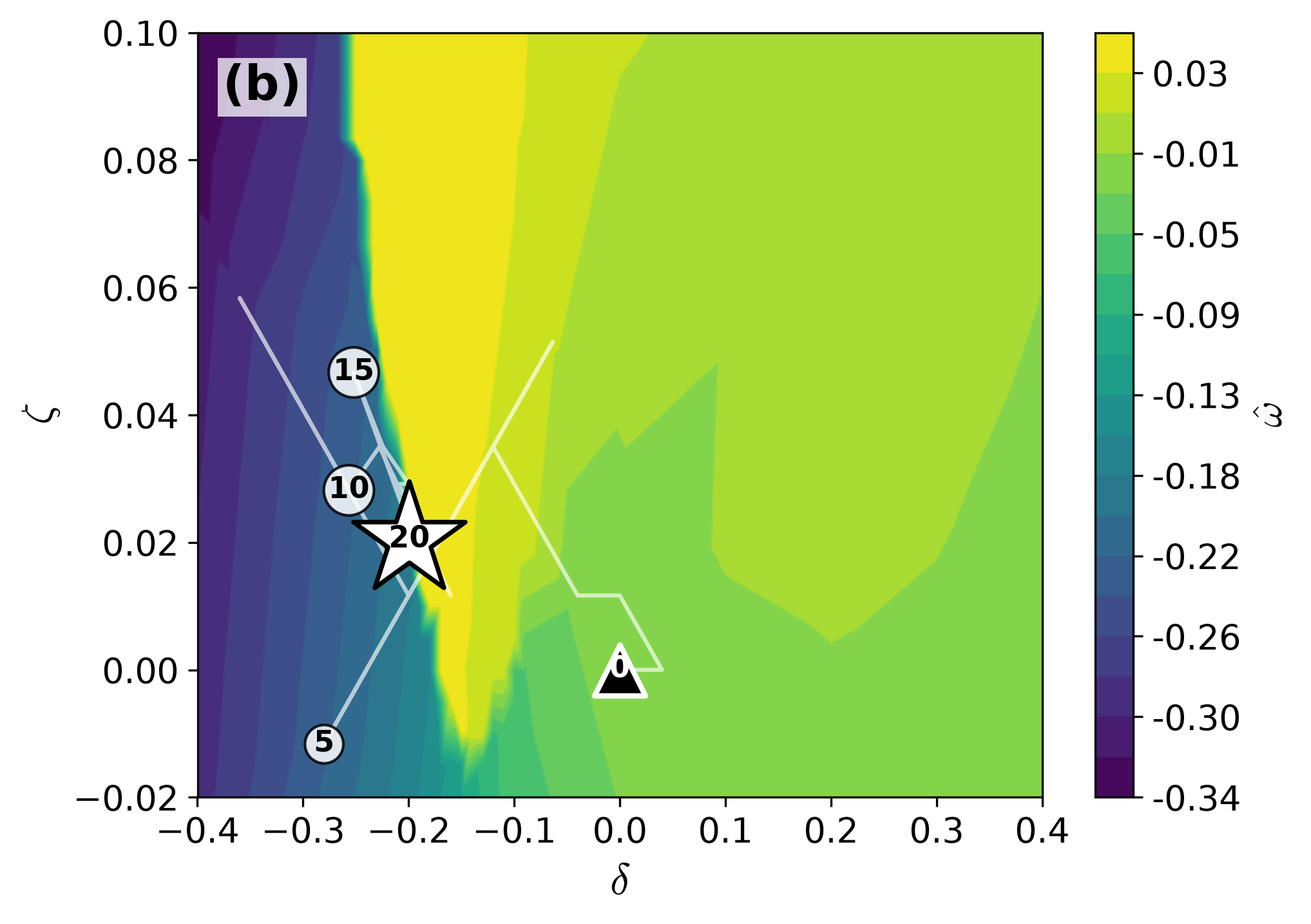}
    \caption{Contour plot over $(\delta,\zeta)$ of
    (a) the objective function $G$ and (b) the corresponding normalized real frequency $\hat{\omega}$, for the case of fixed $s_\delta=s_\zeta=0$. Marker
    overlays indicate the optimization path, starting at $(\delta=0,\zeta=0)$.}
    \label{fig:opt_gamma}
\end{figure}

\begin{figure}
    \centering
    \includegraphics[width=0.49\textwidth]{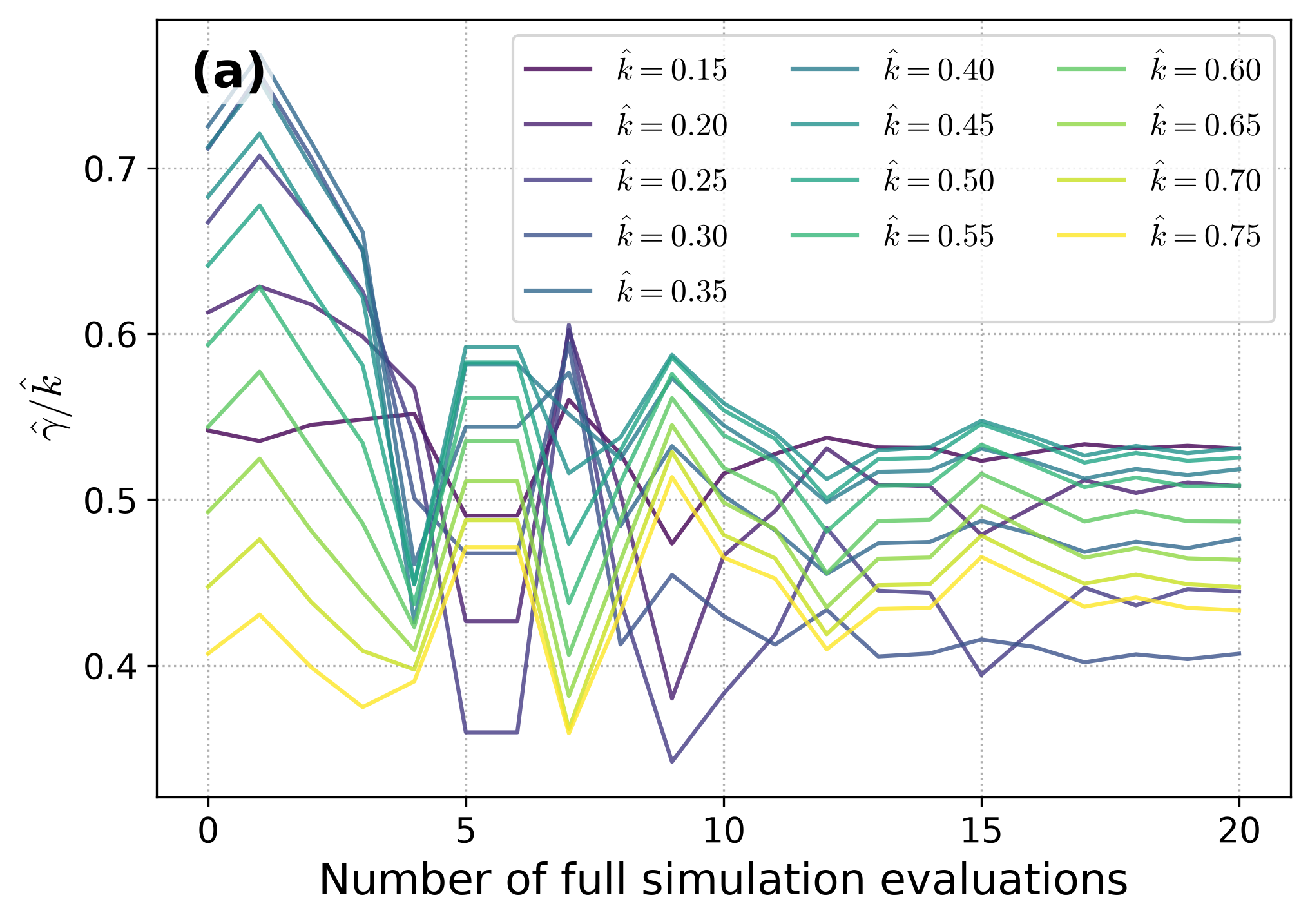}
       \includegraphics[width=0.49\textwidth]{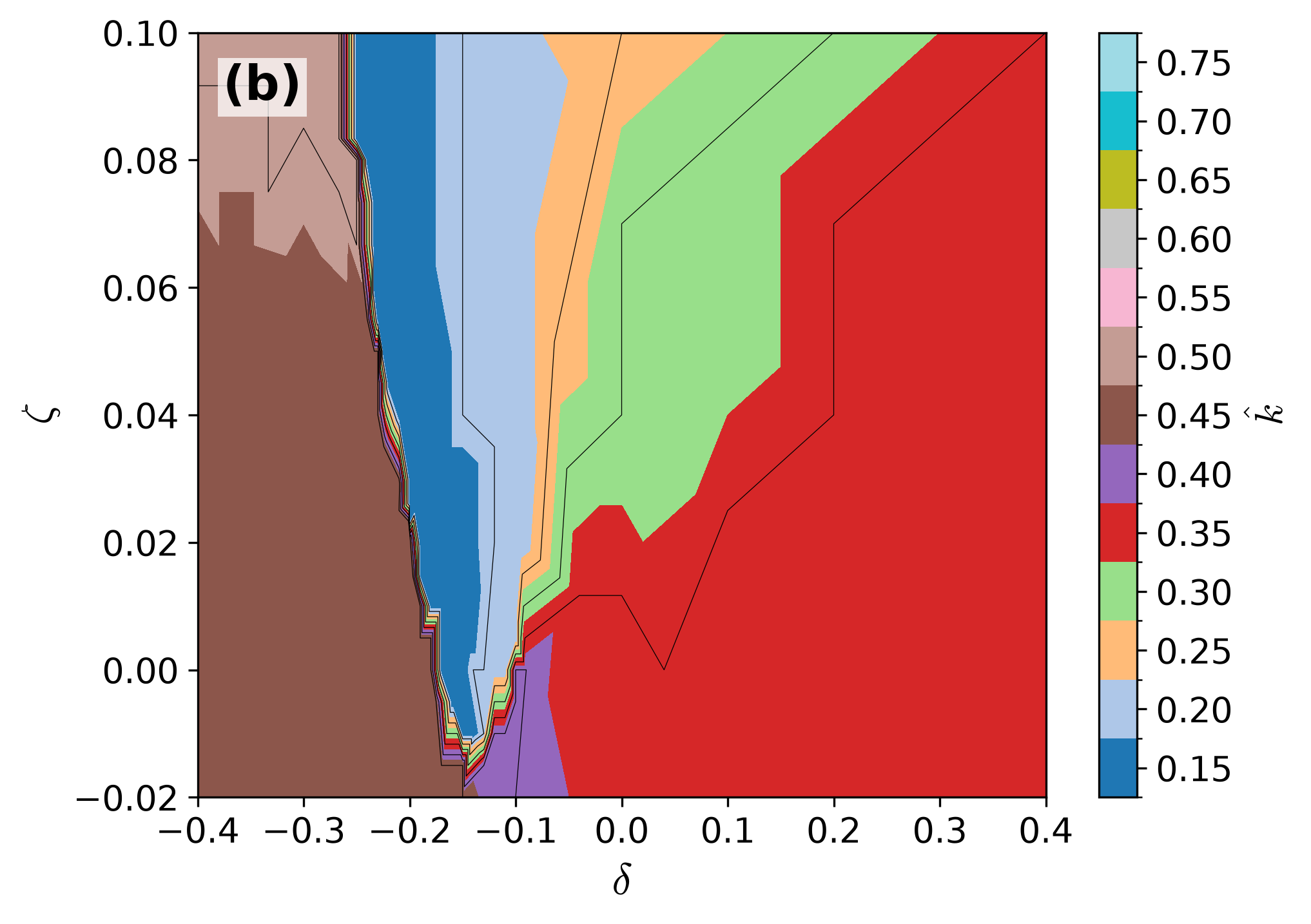}
    \caption{(a) Optimization progress showing the evolution of components of the objective function at each point in the spectrum at each iteration.  The maximizing wavenumber changes over the course of the optimization. (b) Discrete contour plot indicating, for each $(\delta,\zeta)$, the
    value of $\hat{k}$ that maximizes $\hat{\gamma}/\hat{k}$. The colormap is restricted
    to the finite set of sampled $\hat{k}$ values.} 
    \label{fig:opt_k}
\end{figure}

The results of the shaping optimization for the DIII-D baseline case are shown in Fig.~\ref{fig:opt_gamma}a, where $G(x_n)$ denotes the objective function defined in Eq.~(\ref{eqn_obj}), i.e. the maximum of $\hat{\gamma}/\hat{k}$ over the sampled wavenumber spectrum for a set of shaping parameters $x_n$.
For simplicity, we consider only a 2D optimization with $(\delta,\zeta)$ (and their radial derivatives), as explained in Sec.~\ref{sec.shaping}.  We first consider the case with $s_\delta=s_\zeta=0$.  The algorithm is started arbitrarily at $(\delta=0,\zeta=0)$ and the optimal value that minimizes the maximum growth rate-to-wavenumber ratio over ion-scale wavenumbers occurs in the NT region at $(\delta,\zeta)=(-0.199,0.0204)$
after 20 iterations.  Fig.~\ref{fig:opt_k}a shows the corresponding progression of the IBCDFO algorithm.  The optimization history shows the progress of the model-based search, as well as which spectral component is active at each evaluated configuration, consistent with the discrete map in Fig.~\ref{fig:opt_k}b.  The component-wise traces of $\hat{\gamma}(x_n,\hat{k}_j)/\hat{k}_j$ show the non-smooth structure of the objective, in that the maximizing wavenumber can rapidly change over the course of the optimization. This non-smooth composite objective motivates our manifold-sampling approach.  

Physically, the contour plot of $G(\delta,\zeta)$ in Fig.~\ref{fig:opt_gamma}a shows a more-stabilized region in the NT region around $\delta$=-0.1 to -0.3 and a more-destabilized region in the PT region where the objective function increases monotonically with increasing $\delta$.  There is weak sensitivity to $\zeta$ at a given $\delta$.  Fig.~\ref{fig:opt_k}b shows that the value of $k_\theta$ where $(\hat{\gamma}/\hat{k})$ is maximum for a given set of parameters shifts to lower wavenumbers near the optimal $\delta$ to more intermediary values on either side beyond this region.  Near the optimal point in the NT region, there is an abrupt jump (spectral shift) in the location of optimal wavenumber from low to high wavenumbers as $\left| \delta \right|$ increases.  The results in Fig.~\ref{fig:opt_gamma}b also show that the normalized real frequency
$\hat{\omega} = \omega \; (a/c_s)$ at the minimum objective function is positive across the span of the PT region, but shifts to negative as $|\delta|$ increases deep in the NT region across this transitionary region.   In our notation, $\hat{\omega} > 0$ is the ion diamagnetic direction.  Thus, this is consistent with a mode transition from an ion temperature gradient (ITG) mode in the PT region to a dominant trapped electron mode (TEM) in the deep NT region.

\begin{figure}
    \centering
    \includegraphics[width=0.49\textwidth]{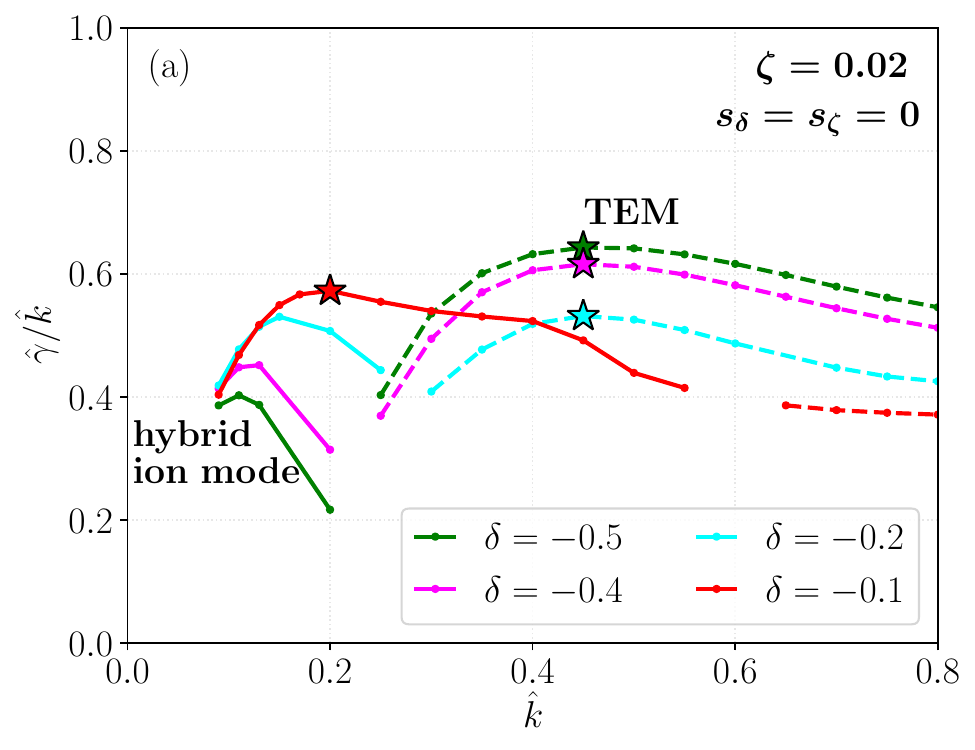}
    \includegraphics[width=0.49\textwidth]{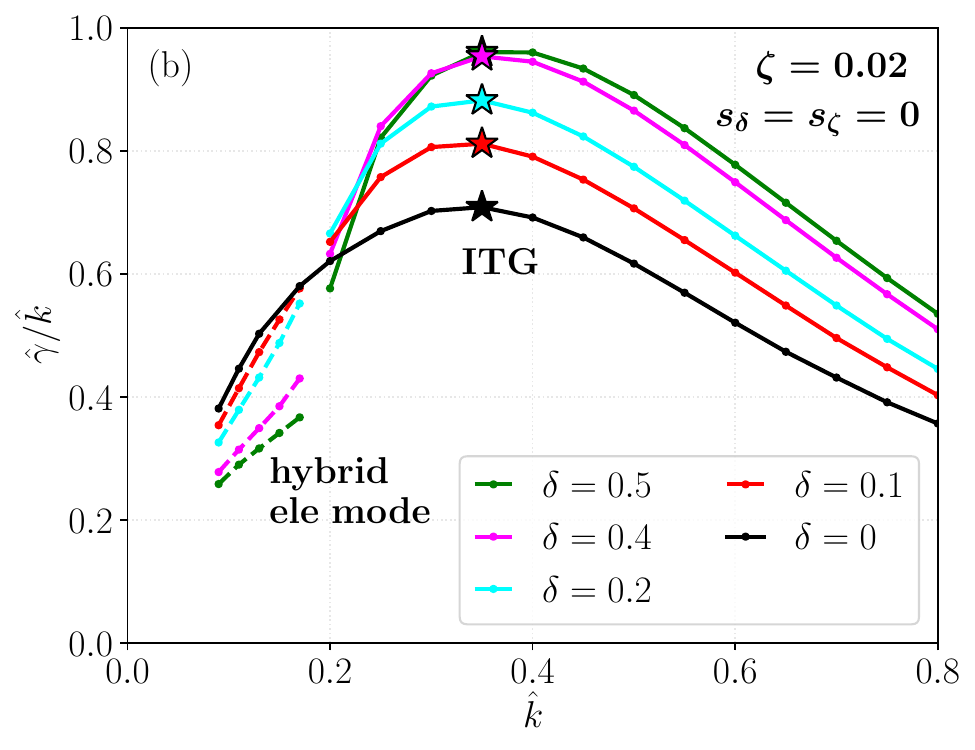}
     \includegraphics[width=0.49\textwidth]{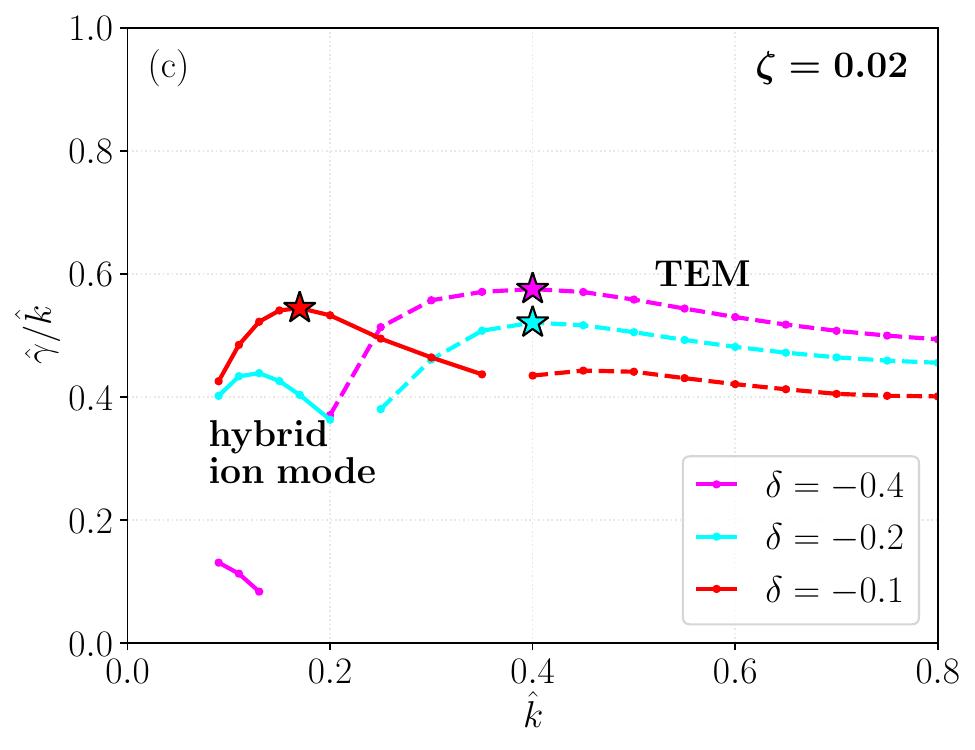}
    \includegraphics[width=0.49\textwidth]{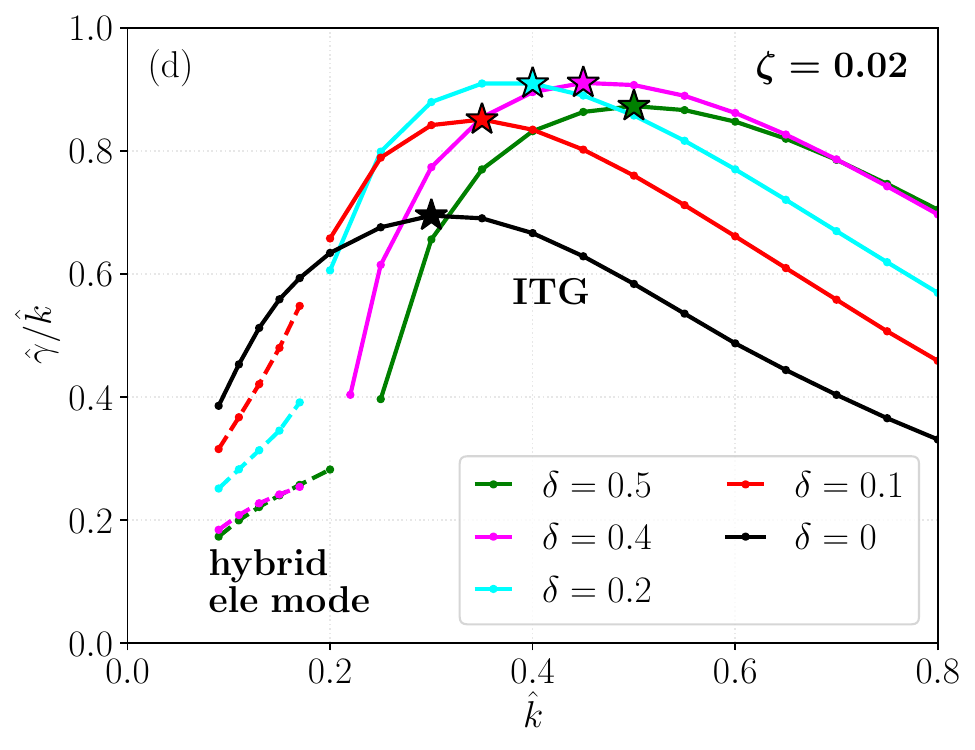}
    \caption{Linear growth rate vs. normalized binormal wavenumber at fixed $\zeta=0.02$ for different values of (a,c) negative triangularity and (b,d) positive triangularity, with fixed $s_\delta=s_\zeta=0$ and varying $s_\delta(\delta)$ and $s_\zeta(\zeta)$.  Stars denote the objective function value $G$ for a given set of shaping parameters, which shifts from a TEM to an ITG mode between the NT and PT regions.}
    \label{fig:lin_gamma}
\end{figure}

To understand the spectral shift better, Fig.~\ref{fig:lin_gamma}(a,b) shows the linear growth rate spectra for various values of $\delta$ at fixed $\zeta=0.02$.  In both regimes, the spectra are complex, with multiple dominant linear instabilities.  The elliptical and PT cases in Fig.~\ref{fig:lin_gamma}b  show the dominant ITG mode for which $\hat{\gamma}/\hat{k}$ increases with increasing $\delta$.  The optimal point in this region is thus at $\delta=0$.  At finite $\delta$, a distinct, yet weakly unstable, hybrid electron mode emerges at low wavenumber.  In contrast, for the NT case in Fig.~\ref{fig:lin_gamma}a, a hybrid ion mode is dominant at low wavenumbers, for which $\hat{\gamma}/\hat{k}$ decreases with increasing $\left| \delta \right|$.  A TEM at higher wavenumbers instead increases with increasing $\left| \delta \right|$. The competition between these modes corresponds to the abrupt transition in the real frequency in Fig.~\ref{fig:opt_gamma}b.  The main point is that the optimization algorithm is able to efficiently handle these non-smooth transitions due to complex mode jumps.

\begin{figure}
    \centering
    \includegraphics[width=0.49\textwidth]{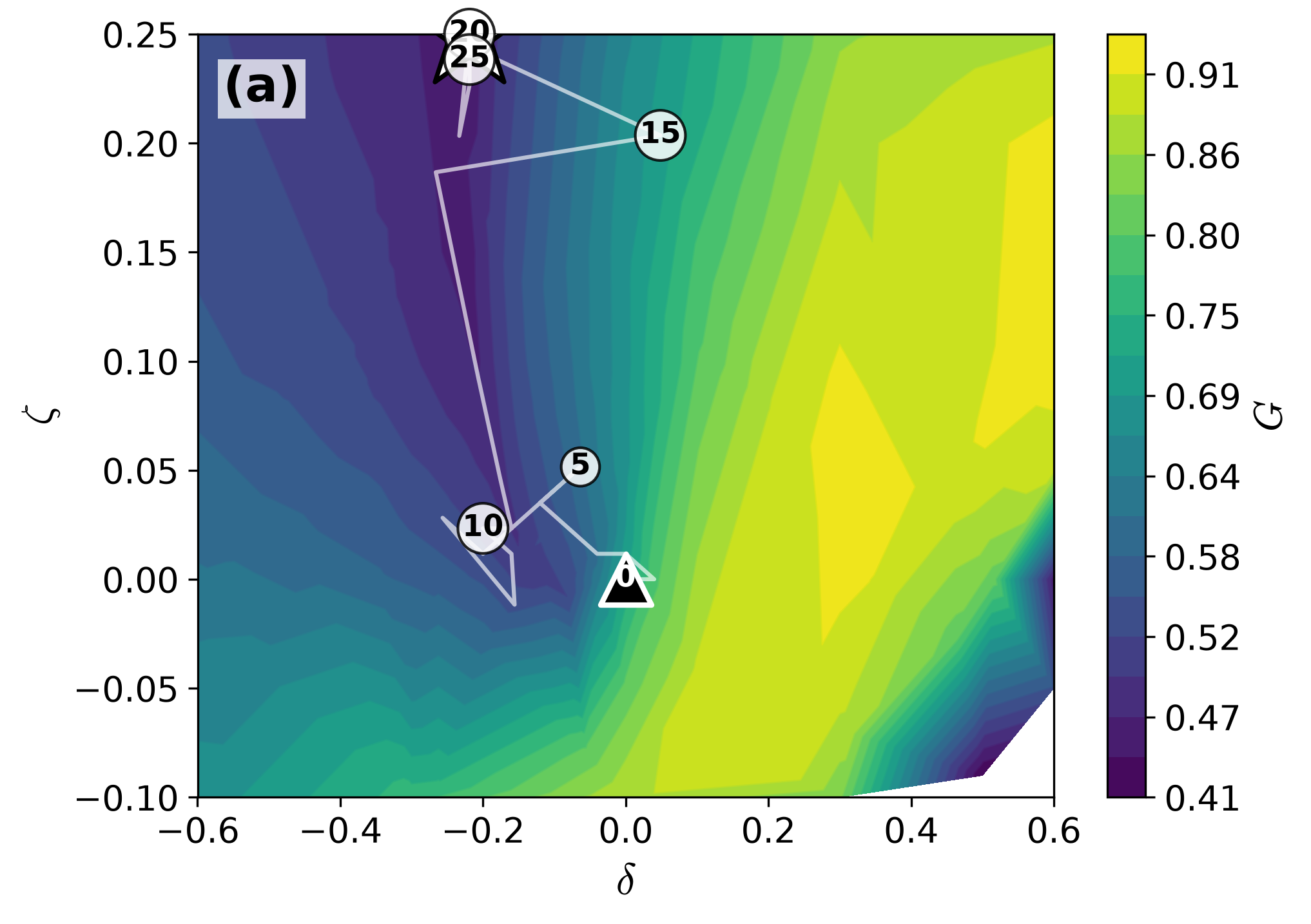}
       \includegraphics[width=0.49\textwidth]{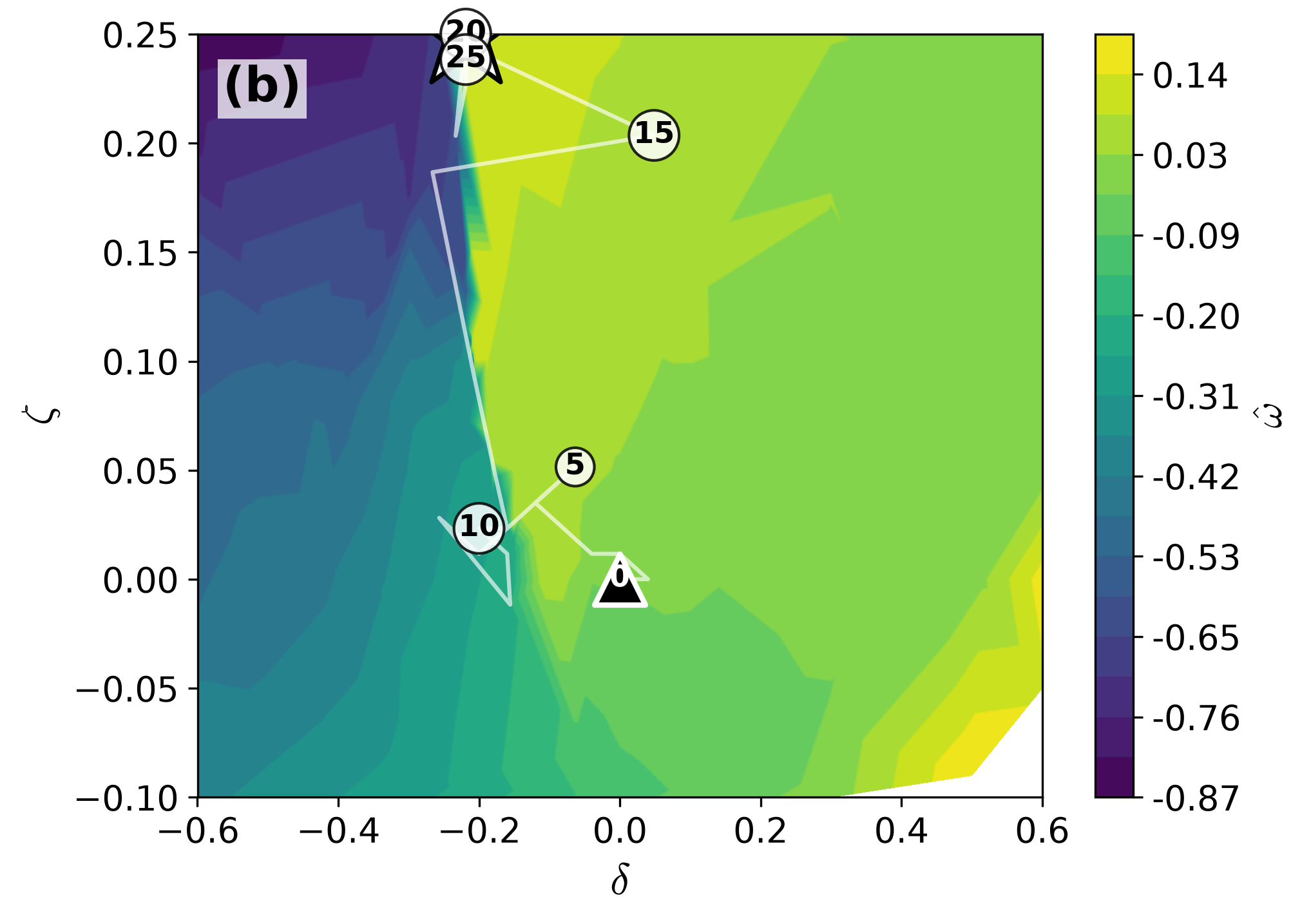}
    \caption{Contour plot over $(\delta,\zeta)$ of
    (a) the objective function $G$ and (b) the corresponding normalized real frequency $\hat{\omega}$, for the case of varying $s_\delta(\delta)$ 
    $ s_\zeta(\zeta)$. Marker
    overlays indicate the optimization path, starting at $(\delta=0,\zeta=0)$.}
    \label{fig:opt_gamma_variable}
\end{figure}

To further demonstrate the robustness of the algorithm, Fig.~\ref{fig:opt_gamma_variable} shows the results for varying $s_\delta(\delta)$ and $s_\zeta(\zeta)$, as described by Eqs.~(\ref{eqn_sdelta}) and (\ref{eqn_szeta}) respectively.  Compared to Fig.~\ref{fig:opt_gamma}, the landscape is much more complex, with multiple regions of mode transitions in both $\delta$ (at fixed $\zeta$) and $\zeta$ (at fixed $\delta$) and multiple local minima and maxima.  The algorithm is started arbitrarily at $(\delta=0,\zeta=0)$ and the optimal value that minimizes the maximum growth rate-to-wavenumber ratio over ion-scale wavenumbers occurs again in the NT region near $\delta=-0.2$ ($\delta_{opt}=-0.206$), but at the upper bound of the squareness range at $\zeta=0.25$.  Ion-direction modes are still dominant in the PT region, while electron-direction modes are dominant in the deep NT region, and there is again an abrupt transition between the two region near $\delta=-0.2$.  The corresponding spectra in Fig.~\ref{fig:lin_gamma}(c,d) show that now in the PT region, the objective function $G$ for the ITG mode shifts to higher wavenumber as $\delta$ increases, corresponding to the eventual decrease in objective for $\delta > 0.4$.  Over the scan, $\delta=0$ is still the optimal value in the PT region.  However, at negative $\zeta$, Fig.~\ref{fig:opt_gamma_variable} shows that high $\delta$ becomes the optimum.  In the NT region, the competition between the TEM and hybrid ion modes is similar to the case of zero shaping gradients.  However, the TEM is more weakly destabilized by $\left|\delta \right|$, while the hybrid ion mode is very strongly stabilized.  Thus, the parameters at which the transition region occurs in Figs.~\ref{fig:opt_gamma}a and \ref{fig:opt_gamma_variable}a are similar.


In terms of shaping optimization, the results are consistent with a benefit of NT vs. PT for gyrokinetic stability and turbulence reduction \cite{marinoni:2009}.  However, for this case we find that, while both NT and (moderate) PT have a positive effect on ion-direction modes, large NT destabilizes the competing TEM branch.  Thus, the optimal shape reflects a trade-off between the two trends.  For squareness, the squareness gradient, rather than the squareness itself, has a larger effect. A limitation of these results is that the present study, like related local gyrokinetic shaping analyses \cite{kinsey:2007,belli:2008b,merlo:2023}, treats the shaping parameters as independently adjustable at a fixed flux surface.  In an experimental or reactor equilibrium, these parameters are constrained by global MHD equilibrium and profile consistency. A fully self-consistent treatment would require coupling the shaping optimization to an evolving global equilibrium and transport solution \cite{highcock:2018}. Extending the present workflow toward such an integrated approach is a natural direction for future work.

\section{Summary and discussion}
\label{sec.summary}

A dynamic optimization workflow for linear gyrokinetic stability has been developed, coupling the gyrokinetic code CGYRO with the libEnsemble ensemble-coordination framework and a manifold-sampling derivative-free optimization algorithm from the IBCDFO package. The workflow is designed to address  practical challenges of simulation-driven optimization in fusion plasma physics, including expensive objective evaluations, unavailable gradient information, and non-smoothness arising from transitions between dominant instability branches. By embedding the optimization algorithm directly within the libEnsemble worker-manager model, the workflow maintains high machine utilization while adaptively focusing computational effort on the most promising regions of the parameter space.

Beyond shaping, the workflow developed here is directly applicable to optimization over any subset of the local equilibrium parameter space accessible to gyrokinetic codes, including profile parameters 
(e.g.~temperature, density, rotation). 
In future work, the workflow will be extended to nonlinear transport objectives using either a reduced (quasilinear) model, such as TGLF \cite{staebler:2005}, or full nonlinear gyrokinetic simulations.  Gradient-based adjoint methods and automatic-differentiation approaches applied to nonlinear gyrokinetic turbulence face their own distinct challenge in this regime, as the chaotic dynamics can make the computed gradient unreliable over long integration windows. The difficulty of extending the present workflow to nonlinear simulations is not only the need for large computational resources, though the ability of libEnsemble to coordinate large ensembles of simulations allows for leveraging scalability if large node counts are available, but also the need for strategies to adjust for resolution requirements in shifting complex turbulence regimes and collapse of transport near marginal stability.   

\section*{Data Availability}

The data that support the findings of this article are available from the authors upon reasonable request.

\section*{References}
\bibliographystyle{iopart-num}
\bibliography{scidac}

\section*{Acknowledgments}
This material is based upon work supported by the U.S.~Department of Energy, Office of Science, Office of Advanced Scientific Computing Research and Office of Fusion Energy Sciences, Scientific Discovery through Advanced Computing (SciDAC) Program through the FASTMath Institute and the Frontiers in Leadership Gyrokinetic Simulation Partnership Project under Contract No.~DE-SC0024425.  Computing resources were provided by the National Energy Research Scientific Computing Center, which is an Office of Science User Facility supported under Contract DE-AC02-05CH11231. 
An award of computer time was also provided by the INCITE program.  This research used resources of the Oak Ridge Leadership Computing Facility, which is an Office of Science User Facility supported  under Contract DE-AC05-00OR22725.   This report was prepared as an account of work sponsored by an agency of the United States Government. Neither the United States Government nor any agency thereof, nor any of their employees, makes any warranty, express or implied, or assumes any legal liability or responsibility for the accuracy, completeness, or usefulness of any information, apparatus, product, or process disclosed, or represents that its use would not infringe privately owned rights. Reference herein to any specific commercial product, process, or service by trade name, trademark, manufacturer, or otherwise, does not necessarily constitute or imply its endorsement, recommendation, or favoring by the United States Government or any agency thereof. The views and opinions of authors expressed herein do not necessarily state or reflect those of the United States Government or any agency thereof.

\end{document}